\documentclass[prc,aps,floatfix,twocolumn,groupedaddress,nofootinbib,showpacs,preprintnumbers,superscriptaddress,
amsmath,amssymb,amsfonts,widetable]{revtex4-2}
\usepackage{array}
\usepackage{hyperref}
\usepackage[nameinlink,capitalize]{cleveref}
\usepackage{xcolor}
\usepackage{graphicx}
\usepackage{bm}
\usepackage{soul}
\usepackage{array}
\usepackage{lipsum}
\usepackage{relsize}
\usepackage{mathrsfs}
\usepackage{mathtools}
\usepackage{amssymb}
\usepackage{amsmath}
\usepackage{graphicx}
\usepackage{microtype}
\usepackage{slashed}
\usepackage[compat=1.1.0]{tikz-feynman}
\usetikzlibrary{positioning}
\usepackage{subcaption} 

\crefname{section}{Sec.}{Secs.}

\preprint{LA-UR-26-22188}

\begin{document}

\title{Electroweak Radiative Corrections to Parity-Violating Electron-Nucleus Scattering}

\author{Brendan T. Reed}
\affiliation{
Theoretical Division, Los Alamos National Laboratory, Los Alamos, NM 87545, USA
}
\email{breed@lanl.gov}

\author{C.J. Horowitz}
\affiliation{Indiana University, Bloomington, IN 47408, USA}
\email{horowit@iu.edu}

\begin{abstract}
Parity-violating electron-scattering provides a largely model independent way of measuring neutron densities in nuclei that has important implications for the structure of nuclei and neutron stars.  In this paper we calculate radiative corrections to the parity-violating asymmetry $A_{\rm pv}$ in electron-nucleus scattering including vertex and vacuum polarization contributions.  We find large cancellations between the vertex corrections to the vector and axial-vector vertices.  As a result the total radiative correction in 2nd Born approximation is dominated by vacuum polarization and is of order -0.5\%.  Coulomb distortions have modest effects on radiative corrections for $^{12}$C and $^{48}$Ca nuclei.  For $^{208}$Pb Coulomb distortions reduce the total radiative correction to only about 0.1\%.  Therefore, these corrections are not important for the interpretation of the PREX and MREX experiments on $^{208}$Pb and for the CREX experiment on $^{48}$Ca. However, radiative corrections must be carefully included for a precision measurement of the weak charge of $^{12}$C.

\end{abstract}

\maketitle

\section{Introduction}
Parity-violating electron-nucleus scattering (PVES) provides a model independent way to probe neutron densities that is free from most strong interaction uncertainties \cite{DONNELLY1989589,PhysRevC.63.025501}. 
In nuclei, one may infer the point neutron radius by measuring the parity-violating asymmetry, $A_{\rm pv}$, from PVES experiments at low momentum transfer.
The neutron radii of $^{208}$Pb, $^{48}$Ca and $^{27}$Al have been measured in the PREX \cite{PhysRevLett.108.112502,PhysRevLett.126.172502}, CREX \cite{PhysRevLett.129.042501} and Qweak \cite{PhysRevLett.128.132501} experiments.  
These radii have implications for nuclear structure \cite{Thiel_2019}, the saturation density of nuclear matter \cite{PhysRevC.102.044321}, the equation of state of neutron-rich matter \cite{PhysRevLett.126.172503}, and the structure of neutron stars \cite{PhysRevLett.86.5647,PhysRevLett.95.122501}. 

\begin{figure*}[htb]
    \centering
    \includegraphics[width=0.9\linewidth]{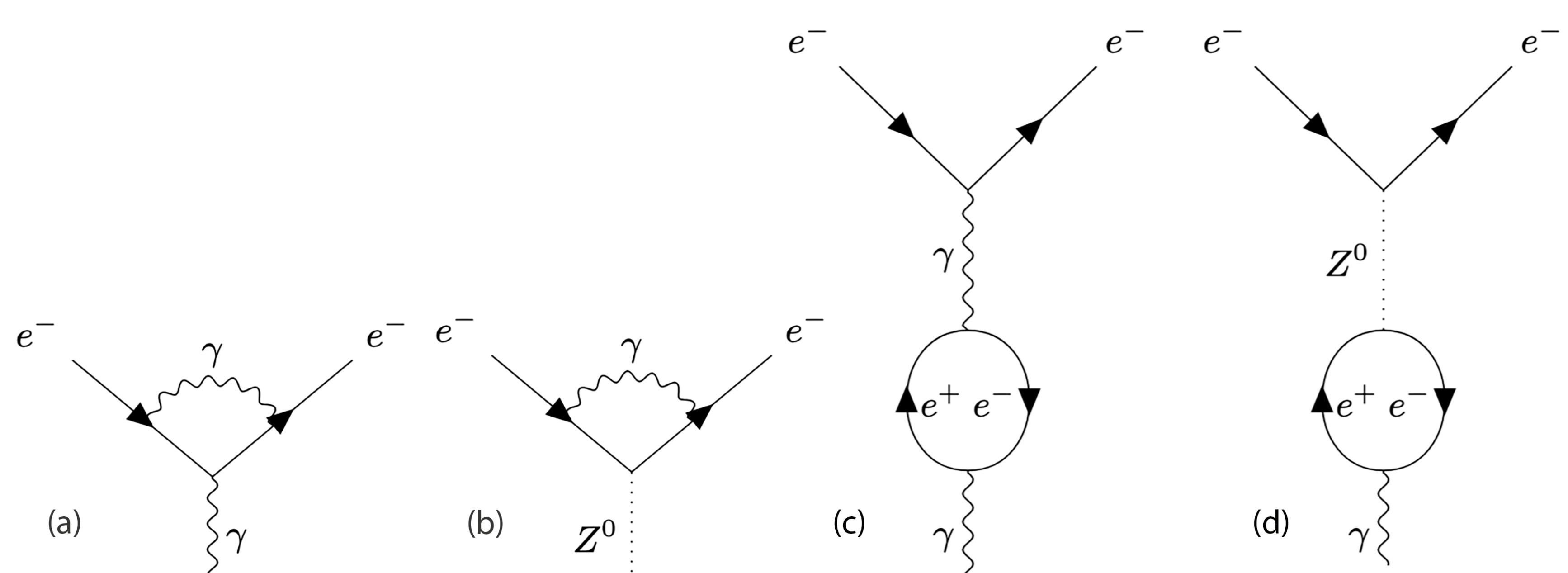}
    \caption{Radiative corrections for the electron leg. Individually shown are the electron vertex diagram for vector $\gamma^\mu$ (a) and for axial-vector $\gamma^\mu\gamma^5$ interactions (b) and vacuum polarization diagrams for the photon (c) and $Z^0$ (d).}
    \label{fig:4diagrams}
\end{figure*}

Recently, Roca-Maza and Jakubassa-Amundsen have claimed a large $\approx 5\%$ radiative correction to $A_{pv}$ in elastic electron-nucleus scattering \cite{PhysRevLett.134.192501}.  
If true, this correction impacts the interpretation of the PREX and CREX experiments and would be even more important for the future MREX \cite{schlimme2024mesaphysicsprogram} and $^{12}$C \cite{PhysRevC.102.022501} experiments. 
For example, the $^{12}$C experiment aims to determine the weak charge of $^{12}$C to 0.3\% and would be sensitive to radiative corrections of order 0.3\% or larger.  
However, Roca-Maza and Jakubassa-Amundsen only consider radiative corrections to the vector interaction shown in \cref{fig:4diagrams} (a) and (c), neglecting the weak vertex radiative correction shown in \cref{fig:4diagrams} (b).  
Therefore, their claim of 5\% radiative corrections is incomplete.  

In this paper, we document {\it both vector and axial-vector} radiative corrections to the electron arm for PVES including all of the diagrams in \cref{fig:4diagrams}. 
There are important cancellations between \cref{fig:4diagrams} (a) and (b) so that the total radiative correction is much smaller than 5\%.  
We begin with the first and 2nd Born approximation in \cref{sec:born} and review vector radiative corrections in \cref{sec:EM}.  
Next we present the first calculations of the axial-vector vertex plus self-energy \cref{fig:4diagrams} (b) in \cref{sec:vertex} and vacuum polarization \cref{fig:4diagrams} (d) in \cref{sec:Zvac}.  
\cref{sec:2ndborn} gives the parity-violating asymmetry in the 2nd Born approximation.  
Corrections to the 2nd Born approximation from Coulomb distortions are discussed in \cref{sec:coulomb}.  \cref{sec:results} presents results using kinematics from the PREX \cite{PhysRevLett.126.172502}, CREX \cite{PhysRevLett.129.042501}, MREX \cite{schlimme2024mesaphysicsprogram}, and $^{12}$C experiments. Finally, we conclude in \cref{sec:conclusions}.



\section{First and 2nd Born Approximations}
\label{sec:born}

\begin{figure}[htb]
\centering
\begin{subfigure}[htb]{0.48\columnwidth}
\centering
\begin{tikzpicture}[scale=0.9]
  \begin{feynman}
    \vertex (a); 
    \vertex (i1) [above left=1.0cm and 1.2cm of a] {\(e^-\)};
    \vertex (f1) [above right=1.0cm and 1.2cm of a] {\(e^-\)};

    \vertex (bn) [below=1.1cm of a];

    \diagram*{
      (i1) -- [fermion] (a) -- [fermion] (f1), 

      (a) -- [photon, edge label'=\(\gamma\)] (bn),
    };
  \end{feynman}
\end{tikzpicture}
\caption*{(a) One photon exchange.}
\end{subfigure}
\hfill
\begin{subfigure}[htb]{0.48\columnwidth}
\centering
\begin{tikzpicture}[scale=0.9]
  \begin{feynman}
    \vertex (a);
    \vertex (i1) [above left=1.0cm and 1.2cm of a] {\(e^-\)};
    \vertex (f1) [above right=1.0cm and 1.2cm of a] {\(e^-\)};

    \vertex (bn) [below=1.1cm of a];

    \diagram*{
      (i1) -- [fermion] (a) -- [fermion] (f1), 

      (a) -- [dotted, edge label'=\(Z^0\)] (bn),
    };

  \end{feynman}
\end{tikzpicture}
\caption*{(b) $Z^0$-boson exchange.}
\end{subfigure}
\hfill
\caption{Tee-level diagrams for parity-violating electron-scattering.}
\label{fig:tree}
\end{figure}
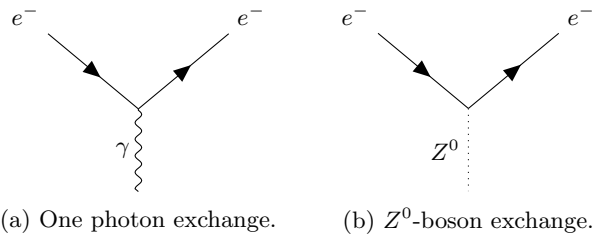

To begin our discussion of radiative corrections, we start with the tree-level Born calculation of the parity-violating asymmetry, $A^b_{\rm pv}$, for electrons scattering from a spin zero nucleus with 4-momentum transfer $-q^2=Q^2>0$.
The Born amplitudes for the Coulomb and weak (axial-vector) interactions (see \cref{fig:tree}) are given by
\begin{eqnarray}
    \bar{u}\gamma^0u\, \mathcal{M}_c=\bar{u}(\bm{k}',h')\gamma^0u(\bm{k},h)\frac{4\pi\alpha}{Q^2}ZF_{ch}(Q^2) \\
    \bar{u}\gamma^0\gamma^5u\,\mathcal{M}_a=\bar{u}(\bm{k}',h')\gamma^0\gamma^5u(\bm{k},h)\frac{G_F}{2^{3/2}}Q_{wk}F_{wk}(Q^2)
\label{eq:tree_amp}
\end{eqnarray}
where $u(\bm{k},h)$ is an electron Dirac-spinor of initial (final) momentum $\bm{k} (\bm{k'})$ and helicity $h$.  
The form factors $F_{ch}$ and $F_{wk}$ are the Fourier transform of the weak and charge densities, respectively.
The total weak charge of the nucleus is $Q_{wk}$ while $Z$ is the nuclear charge. 
We normalize the form factors so that $F_{ch}(0)=F_{wk}(0)=1$.

One sees that $\gamma^5$ in the weak amplitude introduces a helicity dependence, effectively a $\pm\mathcal{M}_a$ for positive or negative helicity electrons because, in the extreme relativistic limit $k\gg m$ with $m$ the electron mass, one has $\gamma^5u(\bm{k},h)\approx hu(\bm{k},h)$.
At the cross-section level for different helicity electrons, we see the rise of an interference term between these two amplitudes,
\begin{eqnarray}
    \frac{d\sigma_{\pm}}{d\Omega}\propto |\mathcal{M}_c|^2\pm2\Re(\mathcal{M}_{c} \mathcal{M}_{a}^*)+\mathcal{O}(G_F^2) 
\end{eqnarray}
where we only keep terms of $\mathcal{O}(G_F)$. 
The parity-violating asymmetry arises from this interference, explicitly given by
\begin{eqnarray}
    A_{\rm pv}^b\equiv \frac{d\sigma_+/d\Omega-d\sigma_-/d\Omega}{d\sigma_+/d\Omega+d\sigma_-/d\Omega} =\frac{G_F Q^2}{4\pi\alpha\sqrt{2}}\frac{|Q_{wk}|}{Z}\frac{F_{wk}(Q^2)}{F_{ch}(Q^2)}
\end{eqnarray}
in the Born approximation.

It is important to note that this form does not contain radiative corrections, which, in the 2nd Born approximation, act like perturbations to the tree-level matrix elements.
If we add for both the Coulomb and axial-vector amplitudes radiative corrections, denoted by $v_v(Q^2)$ and $v_a(Q^2)$ for the Coulomb and axial-vector parts, respectively, then the new matrix elements are simply
\begin{equation}
    \mathcal{M}_{tot} = \mathcal{M}_c(1+v_v(Q^2))\pm\mathcal{M}_a(1+v_a(Q^2))).
\end{equation}
From here we can derive a similar form for the parity-violating asymmetry in the 2nd Born approximation,
\begin{eqnarray}
    A^{(2)}_{pv}=A_{pv}^b\Bigg(\frac{1+v_a(Q^2)}{1+v_v(Q^2)}\Bigg)+\mathcal{O}(A_{pv}^b\alpha^2).
\end{eqnarray}
The Coulombic radiative correction $v_v=v_{vs}+v_\pi$ has contributions from vertex plus self-energy  $v_{vs}$, see \cref{fig:4diagrams}(a), and vacuum polarization $v_\pi$, \cref{fig:4diagrams}(c).  
Milstein and Sushkov calculated vacuum polarization \cite{milstein:2005} while more recently Roca-Maza and Jakubassa-Amundsen find $v_{vs}$ to be large $\approx -0.05$ \cite{PhysRevLett.134.192501}. 
Similarly, the axial radiative correction $v_a=v_{vs}^a+v_\pi^a$ has vertex plus self-energy $v_{vs}^a$ and vacuum polarization $v_\pi^a$ contributions.

\section{Electromagnetic Corrections}
\label{sec:EM}
We first review electromagnetic radiative corrections in this section and then present new results for weak radiative corrections in \cref{sec:weak corrections}.
\subsection{Vacuum Polarization}
Diagram (c) in \cref{fig:4diagrams} shows the first order one photon exchange vacuum polarization arising from a leptonic loop.
This diagram can be written in terms of a first-order correction to the photon propagator,
\begin{eqnarray}
D_{\mu\nu}=D_{\mu\nu}^0+D_{\mu\lambda}^0\Pi^{\lambda\sigma}D_{\sigma\nu}^0,
\end{eqnarray}
where the tree-level propagator is $D^0_{\mu\nu}$ and $\Pi^{\lambda\sigma}$ is the polarization tensor, given by the loop integral
\begin{eqnarray}
    \Pi^{\mu\nu}(q^2) = \alpha\int \frac{d^4k}{(2\pi)^4} \frac{{\rm tr}[\,\gamma^\mu(\slashed{k}+m)\gamma^\nu((\slashed{k}+\slashed{q})+m)]}{(k^2-m^2)((k+q)^2-m^2)}.
\end{eqnarray}
The polarization tensor, due to the Ward-Takahashi identity, can be decomposed into $\Pi^{\mu\nu}=(g^{\mu\nu}q^2-q^\mu q^\nu)\Pi(q^2)$, where the scalar function $\Pi(q^2)$ is \cite{qft_book,Vanderhaeghen:2000}
\begin{eqnarray}
    \Pi(q^2)=\frac{\alpha}{3\pi}\Bigg(v^2-\frac{8}{3}+v\frac{3-v^2}{2}\ln\Big(\frac{v+1}{v-1}\Big)\Bigg)
\end{eqnarray}
with $v^2=1-4m^2/q^2$.
Note that this function has been renormalized by subtracting the full one-loop UV-divergent integral evaluated at $q^2=0$ \cite{qft_book}, so as to reproduce the bare electron charge.
A further approximation that may shed light on the underlying $q$-dependence of this radiative correction is in the limit of $Q^2=-q^2\gg m^2$,
\begin{eqnarray}
    \Pi(Q^2)\approx\frac{\alpha}{3\pi}\Big(-\frac{5}{3}+\ln\Big(\frac{Q^2}{m^2}\Big)\Big).
\end{eqnarray}
The 2nd Born amplitude for this interaction is then given by
\begin{eqnarray}
    \mathcal{M}_{c,\pi} = \frac{4\pi\alpha}{Q^2}ZF_{\rm ch}(Q^2)\Pi(Q^2) \equiv \mathcal{M}_cv_\pi(Q^2).
\label{equation:pi}
\end{eqnarray}
From \cref{equation:pi} we read off the radiative correction,
\begin{equation}
v_\pi(Q^2)=\Pi(Q^2).
\end{equation}
As shown here, this polarization correction grows like $\ln (Q^2/m^2)$ and is of the order $\sim1\%$ for PREX/CREX kinematics.
Importantly, this polarization correction \textit{enhances} the Coulomb amplitude, thereby \textit{decreasing} $A_{pv}$.

\subsection{Vertex Correction}
Similarly, the photon vertex correction is shown in diagram (a) in \cref{fig:4diagrams}.
For an initial electron four momentum $k$ and final momentum $k'$, it is related  to the integral \cite{Vanderhaeghen:2000},
\begin{equation}
    \Gamma_v^\mu=-ie^2\int\frac{d^4l}{(2\pi)^4}\frac{\gamma^\lambda (\slashed k'+\slashed l+m)\gamma^\mu(\slashed k+\slashed l+m)\gamma_\lambda}{(l^2-m_\gamma^2)(l^2+2l\cdot k')(l^2+2l\cdot k)}\, .
\label{vector_vertex}
\end{equation}
This integral is both UV and IR divergent, and must be renormalized carefully as a result.  Here $m_\gamma$ is a small photon mass that regulates the IR divergence.
The on-shell vertex can be parameterized in terms of Fermi and Dirac form factors,
\begin{eqnarray}
    \Gamma_v^\mu(k',k)=F(Q^2)\gamma^\mu-G(Q^2)\frac{i\sigma^{\mu\nu}}{2m}q_\nu
\end{eqnarray}
where $F(Q^2)$ is a Dirac form factor and $G(Q^2)$ is a Pauli form factor.
Note that in the ultrarelativistic limit, $G(Q^2\gg m^2)\rightarrow0$ and so we will neglect its contribution to the 2nd Born amplitude.

To remove the UV divergence from $F(Q^2)$, one requires that the total vertex does not change the physical electron charge at $q^2=0$ \cite{Vanderhaeghen:2000}.
One finds the renormalized form factor 
\begin{equation}
\tilde{F}(Q^2)\equiv F(Q^2)-F(0)=v_{vs}(Q^2)+\text{IR},
\end{equation}
where IR denotes the IR-divergent part of the integral (for $Q^2\gg m^2$),
\begin{equation}
\text{IR}=\bigg(\ln\Big(\frac{Q^2}{m^2}\Big)-1\bigg)\ln\Big(\frac{m_\gamma^2}{m^2}\Big)\, ,
\label{eq:ir_divergence}
\end{equation}
and the vertex plus self-energy radiative correction $v_{vs}(Q^2)$ is
\begin{eqnarray}
    v_{vs}(Q^2) && = \nonumber\frac{\alpha}{2\pi}\Bigg(\frac{v^2+1}{4v}\ln\bigg(\frac{v+1}{v-1}\bigg)\ln\bigg(\frac{v^2-1}{4v^2}\bigg)\\
    \nonumber&&
    +\frac{v^2+1}{2v}\Bigg[{\rm Li}\bigg(\frac{v+1}{2v}\bigg)-{\rm Li}\bigg(\frac{v-1}{2v}\bigg)\Bigg]\\
    &&+\frac{2v^2+1}{2v}\ln\bigg(\frac{v+1}{v-1}\bigg)-2 \Bigg),
\end{eqnarray}
 where Li$(x)=-\int_0^xdt \ln(1-t)/t$ is the Spence function.
The IR divergence can be eliminated for electron-nucleus scattering at the cross-section level, by adding in the soft-photon Bremsstrahlung and self-energy contributions to diagram (a) of \cref{fig:4diagrams}.
We regulate the IR divergence with a small but finite photon mass, $m_\gamma$, which adds to the photon propagator in the loop integral.
In doing so, one sees that the IR divergences cancel \textit{exactly} \cite{qft_book}.

Thus, the form of the 2nd Born amplitude for the vertex plus self-energy diagram is
\begin{eqnarray}
    \mathcal{M}_{c,vs}=\frac{4\pi\alpha}{Q^2}ZF_{\rm ch}(Q^2) v_{vs}(Q^2)\equiv \mathcal{M}_c v_{vs}(Q^2)\,.
\end{eqnarray}
In the ultra-relativistic limit $Q^2\gg m$ we find,
\begin{equation}
    v_{vs}(Q^2)=\frac{\alpha}{2\pi}\Bigg(\frac{3}{2}\ln\bigg(\frac{Q^2}{m^2}\bigg)-\frac{1}{2}\ln^2\bigg(\frac{Q^2}{m^2}\bigg) +\frac{\pi^2}{6}-2\Bigg).
\label{eq:vvs}\end{equation}
In this interaction, the correction grows like $\ln^2Q$ which leads to a much larger correction to the Born amplitude than for the vacuum polarization.
In fact, this correction is of the order $-5\%$ for PREX/CREX kinematics, and \textit{decreases} the total Coulomb amplitude.



\section{Radiative Corrections to Weak Interactions}
\label{sec:weak corrections}

We now present new results for the weak radiative corrections from diagrams (b) and (d) in \cref{fig:4diagrams}.  For a spin zero nucleus, parity violation involves an axial coupling of the $Z^0$ to the electron and a vector coupling to the nucleus.  Therefore, we are interested in radiative corrections to the axial-vector interaction.

\subsection{Axial-vector vertex}
\label{sec:vertex}
Here we detail the calculation of the axial-vector vertex correction of \cref{fig:4diagrams}(b).
The loop integral for the axial-vector vertex is, see for example \cite{Vanderhaeghen:2000},
\begin{equation}
    \Gamma_{a}^\mu=-ie^2 \int\frac{d^4l}{(2\pi)^4}\frac{\gamma^\lambda (\slashed k'+\slashed l+m)\gamma^\mu\gamma^5(\slashed k+\slashed l+m)\gamma_\lambda}{(l^2-m_\gamma^2)(l^2+2l\cdot k')(l^2+2l\cdot k)}\,,
\label{eq:Gamma_v}
\end{equation}
with $k$ the initial and $k'$ the final electron momentum and $m$ the electron mass.  
This integral has both IR and UV divergences. 
As before, we introduce a small photon mass $m_\gamma$ which regulates the IR divergence.
The UV divergence is regulated in $D=4-2\epsilon$ dimensions (after moving $\gamma^5$ to the right \cite{PhysRevD.83.025020}). 
The momentum transfer is $q=k^\prime -k$ and we define $Q^2=-q^2>0$.  
Notably, the axial-vector vertex has both $\gamma^\mu\gamma^5$ and pseudoscalar $q^\mu\gamma^5$ contributions,
\begin{equation}
    \Gamma_a^\mu(Q^2)=\Gamma_a(Q^2)\gamma^\mu\gamma^5+\Gamma_{ps}(Q^2)\frac{q^\mu}{m}\gamma^5\, ,
\end{equation}
with
\begin{eqnarray}
\nonumber\Gamma_a&=&\frac{\alpha}{2\pi}\int_0^1 d\xi\int_0^1 dw w \Bigl\{\frac{1}{\epsilon}-\gamma-2+\ln\Big(\frac{4\pi^2\mu^2}{\Delta}\Big)\\
\nonumber&-&\frac{(2-2w+w^2)m^2+(1-w+w^2\xi(1-\xi))Q^2}{\Delta}\Bigr\}\, .
\end{eqnarray}
Here $\gamma(=0.5772...)$ is Euler's constant, $\mu$ is the renormalization scale and $\Delta
= w^2\big(m^2+Q^2\xi(1-\xi)\big) + m_\gamma^2$. 
At $Q^2=0$ the vertex is,
\begin{equation}
\Gamma_a(0)=\frac{\alpha}{2\pi}\Bigl(\frac{1}{2\epsilon}-\frac{\gamma}{2}+\frac{1}{2}\ln\Big(\frac{4\pi^2\mu^2}{m^2}\Big)+\ln\Big(\frac{m_\gamma^2}{m^2}\Big)+1\Bigr)\, ,
\end{equation}
while in the ultra relativistic limit $Q^2\gg m^2$ one has
\begin{eqnarray}
\nonumber\Gamma_a(Q^2)&=&\frac{\alpha}{2\pi}\Bigl(\frac{1}{2\epsilon}-\frac{\gamma}{2}+\frac{1}{2}\ln\Big(\frac{4\pi^2\mu^2}{m^2}\Big)\\\nonumber
&+&\ln\Big(\frac{m_\gamma^2}{m^2}\Big)\ln\Big(\frac{Q^2}{m^2}\Big)\\
&+&\frac{3}{2}\ln\Big(\frac{Q^2}{m^2}\Big)-\frac{1}{2}\ln^2\Big(\frac{Q^2}{m^2}\Big)+\frac{\pi^2}{6}\Bigr)\, .
\end{eqnarray}
As in the E+M case, subtracting $\Gamma_a(0)$ cancels the UV divergence and yields,
\begin{equation}
 \Gamma_a(Q^2)-\Gamma_a(0)= v_{vs}^a+\text{IR}_a,
\end{equation}
with the radiative correction,
\begin{equation}   
v_{vs}^a= \frac{\alpha}{2\pi}\bigg(
\frac{3}{2}\ln\Big(\frac{Q^2}{m^2}\Big)-\frac{1}{2}\ln^2\Big(\frac{Q^2}{m^2}\Big)+\frac{\pi^2}{6}-1\Bigr)\, ,
\label{eq:vavs}
\end{equation}
and the infrared divergent part $\text{IR}_a$ (see also \cref{eq:ir_divergence}),
\begin{equation}
\text{IR}_a=\text{IR}=\bigg(\ln\Big(\frac{Q^2}{m^2}\Big)-1\bigg)\ln\Big(\frac{m_\gamma^2}{m^2}\Big)\, .
\end{equation}
As in the electromagnetic case, one may remove the IR divergence by adding in the soft-photon Bremsstrahlung contribution.

The pseudoscalar vertex is given by,
\begin{equation}
\Gamma_{ps}(Q^2)=\frac{\alpha}{2\pi}\int_0^1 d\xi\int_0^1 dw w \Bigl( \frac{1+w-4w\xi(1-\xi)}{1+\frac{Q^2}{m^2}\xi(1-\xi)}\Bigr)\, .
\end{equation}
This is,
\begin{equation}
\Gamma_{ps}(Q^2)=\frac{\alpha}{2\pi}\frac{m^2}{Q^2}\Bigl[\frac{v^2+2}{v}\ln\Big(\frac{v+1}{v-1}\Big)-2\Bigr]\, ,
\end{equation}
where $v=(1+4m^2/Q^2)^{1/2}$.  As $Q^2\rightarrow 0$,
\begin{equation}
\Gamma_{ps}(Q^2\rightarrow 0)=\frac{\alpha}{2\pi}\frac{7}{6}\, ,
\end{equation}
while,
\begin{equation}
\Gamma_{ps}(Q^2\gg m^2)=\frac{\alpha}{2\pi}\frac{m^2}{Q^2}\Bigl(3\ln\Big(\frac{Q^2}{m^2}\Big)-2\Bigr)\, .
\end{equation}
Often $\Gamma_{ps}$ does not contribute if it couples to a conserved current $\bm{j}$ with $\bm{q}\cdot \bm{j}=0$.
Additionally, this term goes to zero as $q\rightarrow\infty$, so we will not include it in our calculation of the 2nd Born amplitude for the axial-vector vertex.
As a result, the total correction relevant for our study of parity-violating electron-scattering off a \textit{spin-zero} nucleus is given by $v^a_{vs}$ in \cref{eq:vavs}.
The 2nd Born amplitude for this correction is 
\begin{eqnarray}
    \mathcal{M}_{a,vs}=\frac{G_F}{2\sqrt{2}}|Q_{wk}|F_{wk}(Q^2)v_{vs}^a(Q^2)\equiv \mathcal{M}_{a} v_{vs}^a(Q^2).
\end{eqnarray}
As in the E+M case, this correction \textit{reduces} the axial-vector amplitude.

We see that the axial radiative correction $v_{vs}^a$ is very closely related to $v_{vs}$ in \cref{eq:vvs},
\begin{eqnarray}
v^a_{vs}(Q^2)=v_{vs}(Q^2)+\frac{\alpha}{2\pi}\, ,
\label{eq:vavs-vvs}
\end{eqnarray}
as discussed in \cref{sec:discussion}.  Therefore, $v_{vs}$ and $v_{vs}^a$ {\it largely cancel} for $A_{pv}^{(2)}$.

\subsection{Weak Vacuum Polarization}
\label{sec:Zvac}

Similarly as in the photon vacuum polarization, the $Z^0$ polarization shown in diagram (d) of \cref{fig:4diagrams} gives rise to a 2nd Born amplitude,
\begin{eqnarray}
\nonumber    \mathcal{M}_{a,\pi}&=&\frac{G_F}{2\sqrt{2}}ZF_{\rm ch}(Q^2)(1-4\sin^2\theta_W)\Pi(Q^2)\\
    &\equiv&\mathcal{M}_av^a_\pi(Q^2)
\end{eqnarray}
and a radiative correction $v_\pi^a$.  Note the small factor $1-4\sin^2\theta_W$ that corresponds to the weak charge of the electron.  As a result, $v_\pi^a$ is smaller than the other radiative corrections.  This contribution involves only electromagnetic couplings to the nucleus and is independent of the nuclear weak charge $Q_{wk}$.  However since the Born amplitude $\mathcal{M}_a$ includes $Q_{wk}F_{wk}(Q^2)$, these factors are divided out in the definition of $v_\pi^a$,
\begin{eqnarray}
   v^a_\pi(Q^2)= (1-4\sin^2\theta_W)\frac{ZF_{\rm ch}(Q^2)}{|Q_{wk}|F_{wk}(Q^2)}\Pi(Q^2).
\end{eqnarray}

\subsection{Asymmetry in 2nd Born Approximation}
\label{sec:2ndborn}
Now that we have detailed the exact form of each radiative correction, we calculate $A_{\rm pv}$ in 2nd Born approximation, denoted $A_{\rm pv}^{(2)}$, to show the impact of each correction. We have
\begin{eqnarray}
    A_{\rm pv}^{(2)} = A_{\rm pv}^b\Big(\frac{1+v_\pi^a(Q^2)+v_{vs}^a(Q^2)}{1+v_\pi(Q^2)+v_{vs}(Q^2)}\Big)
    \label{eq:2nd_born_Apv}
\end{eqnarray}
where $A_{\rm pv}$ in first Born approximation is denoted $A_{\rm pv}^b$.
Written in this way, one can calculate the radiative correction for each component relative to the tree-level Born asymmetry value.

Here, we can illustrate the cancellation that occurs between the Coulomb and weak vertex corrections.
If we note that for the $Q^2$ values of interest $v_\pi^a\ll v_{vs}^a$ and $v_\pi+v_{vs}\ll 1$, then we may write \cref{eq:2nd_born_Apv} as
\begin{eqnarray}
    A_{\rm pv}^{(2)}\approx A_{\rm pv}^b \big(1+v_{vs}^a-v_\pi-v_{vs}\big)=A_{\rm pv}^b\big(1-v_\pi+\frac{\alpha}{2\pi}\big)\, \, \,
\end{eqnarray}
where the factor of $\alpha/2\pi$ is leftover from subtracting the axial and Coulomb vertex corrections, see \cref{eq:vavs-vvs}.  
Effectively this leaves the only surviving correction as the Coulomb vacuum polarization as all other corrections either cancel or contribute at the $0.1\%$ level in the second Born approximation.
Nonetheless, it is important to also investigate if this holds true beyond the Born approximation.
In the next section, we detail our calculation of $A_{\rm pv}$ with Coulomb distortions.

\section{Coulomb distortions}
\label{sec:coulomb}

We now turn our attention to implementing these corrections in a Coulomb distorted calculation of $A_{\rm pv}$.
We assume that a Dirac equation which is satisfied by parity-violating electron-scattering (PVES) is given by
\begin{eqnarray}
    [\bm{\alpha}\cdot\bm{p}+(V_{\rm ch} + \Delta V_{\rm ch})\pm (A+\Delta A)+\beta m]\Psi=E\Psi
    \label{eq:dirac}
\end{eqnarray}
where $V_{\rm ch}$ is the Coulomb potential, $A$ is the axial vector potential, and terms with $\Delta$ represent radiative corrections.
For the Coulomb and axial-vector potentials, their forms are given by
\begin{eqnarray}
    &&V_{\rm ch}(r) = \int d^3r'\frac{\rho_{ch}(r')}{|\bm{r}-\bm{r}'|}\\
    &&A(r) = \frac{G_F}{2^{3/2}}\rho_{wk}(r).
\end{eqnarray}
The form for each of the radiative corrections are given directly from their second Born matrix elements.
Explicitly, each Coulomb contribution is
\begin{eqnarray}
    &&\Delta V_{\rm ch}=V_\pi+V_{vs},\nonumber\\
    &&V_\pi(r) = \int d^3Q e^{-i(\bm{r}\cdot\bm{Q})}\frac{4\pi\alpha}{Q^2}ZF_{\rm ch}(Q^2)v_\pi,\\
    &&V_{vs}(r) = \int d^3Q e^{-i(\bm{r}\cdot\bm{Q})}\frac{4\pi\alpha}{Q^2}ZF_{\rm ch}(Q^2)v_{vs},
\end{eqnarray}
and, likewise, each axial-vector contribution is
\begin{eqnarray}
    &&\Delta A=A_\pi+A_{vs},\nonumber\\
    &&A_\pi(r) = \int d^3Q e^{-i(\bm{r}\cdot\bm{Q})} \frac{G_F}{2^{3/2}}|Q_{wk}|F_{\rm wk}(Q^2)v_\pi^a,\\
    &&A_{vs}(r) = \int d^3Q e^{-i(\bm{r}\cdot\bm{Q})} \frac{G_F}{2^{3/2}}|Q_{wk}|F_{\rm wk}(Q^2)v^a_{vs}.\, \,
\end{eqnarray}

For our Coulomb-distorted analysis of $A_{\rm pv}$, we calculate the scattering cross-section using the eikonal method.
The scattering amplitude is calculated by evaluating the following integral,
\begin{eqnarray}
    f_\pm(\theta)=-\frac{ik}{2\pi}\int d^2\bm{b} \exp({i\bm{q}\cdot \bm{b}})(\exp(-i\chi_\pm(b))-1)\\
    \chi_\pm(b)=\int_{-\infty}^{\infty}\Big(V_{\rm ch}(r)+\Delta V_{\rm ch}\pm (A(r)+\Delta A)\Big)dz,\, 
\end{eqnarray}
where $r=\sqrt{z^2+b^2}$, $k$ is the 3-momentum of the incident electron and the orthogonal impact parameter $b$ is integrated over in the phase-shift function $\chi_\pm(b)$.
The Born scattering cross-section in the eikonal approximation is given by
\begin{eqnarray}
    \frac{d\sigma_\pm}{d\Omega}=|f_\pm(\theta)|^2\cos^2\frac{\theta}{2},
\end{eqnarray}
therefore the parity-violating asymmetry then is just the difference between solutions with $V_{\rm ch}+A$ and solutions with $V_{\rm ch}-A$:
\begin{eqnarray}
    A_{\rm pv}=\frac{d\sigma_+/d\Omega-d\sigma_-/d\Omega}{d\sigma_+/d\Omega+d\sigma_-/d\Omega}.
\end{eqnarray}

\begin{figure}[thb]
    \centering
    \includegraphics[width=1\columnwidth]{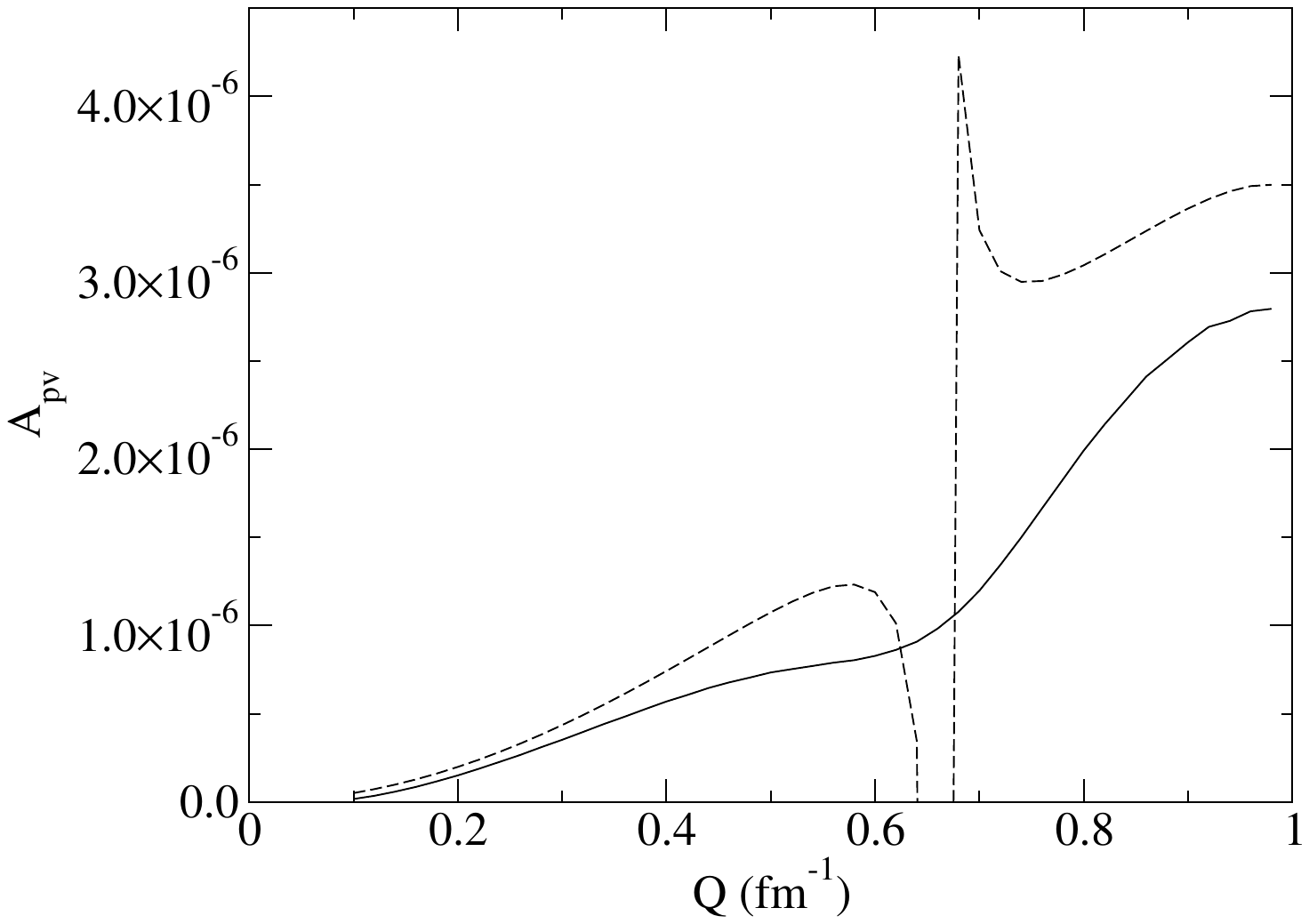}
    \caption{Parity-violating asymmetry $A_{pv}$ for $^{208}$Pb at 1 GeV versus momentum transfer $Q$ calculated in Born approximation (dashed) and with Coulomb distortions (solid).  Neither curve includes radiative corrections.}
    \label{fig:Apv208Pb}
\end{figure}

We note here that the eikonal approximation has been shown to hold well for parity-violating electron-scattering \cite{milstein:2005,dong:2008} and we have confirmed our results are consistent with partial wave analyses \cite{Horowitz:1998vv}. 
\cref{fig:Apv208Pb} shows $A_{pv}$ for $^{208}$Pb. 
In Born approximation $A_{pv}$ has sharp structures in diffraction minima because $F_{ch}(Q^2)$ and $F_{wk}(Q^2)$ have zeros in different locations. 
Coulomb distortions smooth out this structure and significantly reduce $A_{pv}$.

\begin{figure}[htb]
    \centering
    \includegraphics[width=1.0\columnwidth]{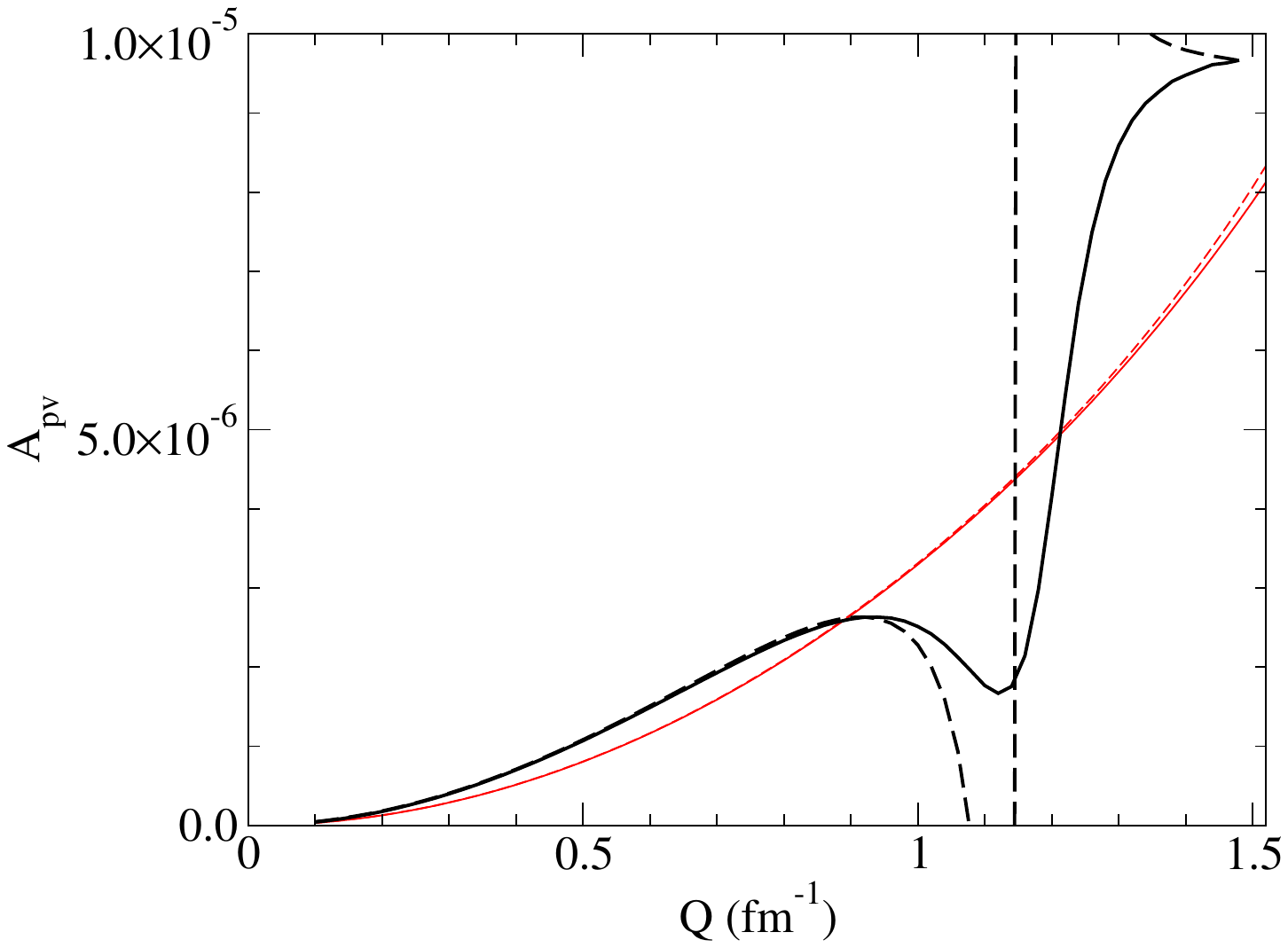}
    \caption{Parity-violating asymmetry $A_{pv}$ for $^{48}$Ca at 2.18 GeV versus momentum transfer $Q$ calculated in Born approximation (heavy black dashed curve) and with Coulomb distortions (heavy black solid curve).  Also shown is $A_{pv}$ for $^{12}$C at 150 MeV (thin red curves).}
    \label{fig:Apv48Ca}
\end{figure}

Coulomb distortions are less important for lighter nuclei.  \cref{fig:Apv48Ca} shows $A_{pv}$ for $^{48}$Ca and for $^{12}$C.  For $^{48}$Ca Coulomb distortions are unimportant away from diffraction minima; however, distortions still have significant effects near diffraction minima.  For $^{12}$C distortion effects are small.

To assess the impact of radiative corrections, we solve the eikonal equations above for different combinations of $V^*_{\rm ch}=V_{\rm ch}+\Delta V$ and $A^*=A+\Delta A$, with and without each radiative correction included.
Because the vacuum polarization of the $Z^0$-boson has been shown to be small for our kinematic range of interest, we choose to only asses the impact of the weak axial-vector vertex correction on the axial-vector potential.

\section{Results}
\label{sec:results}
Here we detail our results beginning with 2nd Born approximation results.
For the nuclear charge and weak densities, we utilize a 2-parameter symmetrized Fermi function \cite{Piekarewicz:2016vbn,Reed_2020st} with parameters found in \cref{tab:fermi_params} and \cref{tab:fermi_params_carbon}.
This parameterization provides an analytic expression for both the density 
\begin{eqnarray}
    &&\rho(r;a,c) = \frac{\rho_0\sinh(c/a)}{\cosh(r/a)+\cosh(c/a)} \\
    &&\nonumber\rho_0=\frac{3\mathcal{Z}}{(4\pi c(c^2+\pi^2a^2))},
\end{eqnarray}
and form factor
\begin{eqnarray}
    && F(Q;a,c) = \frac{Q\pi a F_0}{\tanh(Q\pi a)\sin(Qc)-Qc\cos(Qc))} \\
    &&\nonumber F_0 = \frac{3\pi a/\sinh(Q\pi a)}{Q^2c(c^2+(\pi a)^2)},
\end{eqnarray}
in terms of the nuclear surface thickness ($a$) and radius ($c$),
where $\mathcal{Z}$ is the nuclear charge ($Z$ or $Q_{wk}$ here).
Note that the form factor is normalized $F(Q=0)=1$.
The choice of this simple parameterization aids in providing a model-agnostic density profile and greatly simplifies the calculation of radiative correction potentials and their respective densities, given by
\begin{subequations}
\label{eq:densities}
\begin{align}
     &\rho_{vs}(r) = \frac{Z}{2\pi}\int dQ j_0(Qr)F_{\rm ch}(Q^2)v_{vs}(Q^2),\\
     &\rho_{\pi}(r) = \frac{Z}{2\pi}\int dQ j_0(Qr)F_{\rm ch}(Q^2)v_\pi(Q^2),\\
     &\rho_{vs}^a(r) = \frac{Q_{\rm wk}}{2\pi}\int dQ j_0(Qr)F_{\rm wk}(Q^2)v^a_{vs}(Q^2),\\
     &\rho_{\pi}^a(r) = \frac{Q_{\rm wk}}{2\pi}\int dQ j_0(Qr)F_{\rm wk}(Q^2)v^a_\pi(Q^2).
\end{align}
\end{subequations}
We show these densities for $^{208}$Pb and $^{48}$Ca in \cref{fig:density} and for $^{12}$C in \cref{fig:carbon_densities}.
The radiative corrections presented below, however, are nearly independent of the choice of nuclear densities.
\begin{table}[htbp]
\begin{center} 
  \begin{tabular}{|l | c c c | c c c | }
    \hline
     & & $^{208}$Pb  & & & $^{48}$Ca &  \\
 Density    &$c($fm)  & $a($fm) & $Q_W$ & $c($fm) & $a($fm) & $Q_W$ \\
    \hline
Charge & 6.666 & 0.5122 & & 3.706 & 0.5245 & \\
Weak & 6.815 & 0.6140 & -118.55 & 3.996 & 0.5154 & -26.2164 \\ \hline
    \end{tabular}
    \end{center}\caption{Fermi function radius $c$, surface thickness $a$, and weak charge $Q_W$ parameters for charge $\rho_{ch}(r)$ and weak $\rho_W(r)$ densities for Lead, and Calcium nuclei.} \label{tab:fermi_params} 
\end{table}

\begin{table}[htbp]
\begin{center} 
  \begin{tabular}{|l | c c c |}
    \hline
 & & $^{12}$C &\\
 Density    &$c($fm)  & $a($fm) & $Q_W$ \\
    \hline
Charge & 2.133 & 0.494 & \\
Weak & 2.070 & 0.494 & -5.4942  \\ \hline
    \end{tabular}
    \end{center}\caption{Fermi function radius $c$, surface thickness $a$, and weak charge $Q_W$ parameters for charge $\rho_{ch}(r)$ and weak $\rho_W(r)$ densities for Carbon.} \label{tab:fermi_params_carbon} 
\end{table}
\begin{figure*}[thb]
    \centering
    \includegraphics[width=.9\linewidth]{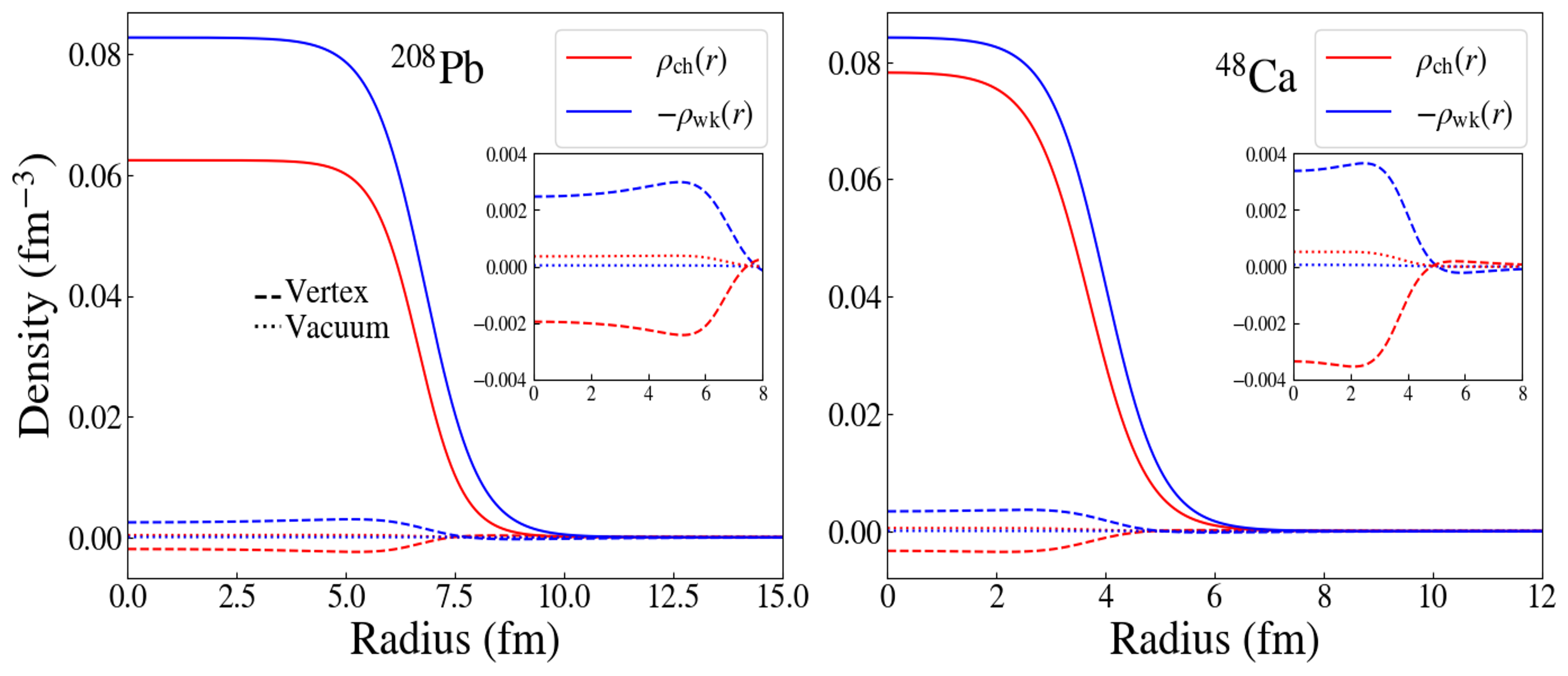}
    \caption{Plot of charge (red) and weak (blue) densities in $^{208}$Pb (left) and $^{48}$Ca (right). Vacuum polarization (dotted) and vertex (dashed) corrections are also shown for each density according to \cref{eq:densities}. The weak densities are multiplied by -1 to show their effective scale compared to the charge densities.}
    \label{fig:density}
\end{figure*}

\begin{figure}[htb]
    \centering
    \includegraphics[width=1\linewidth]{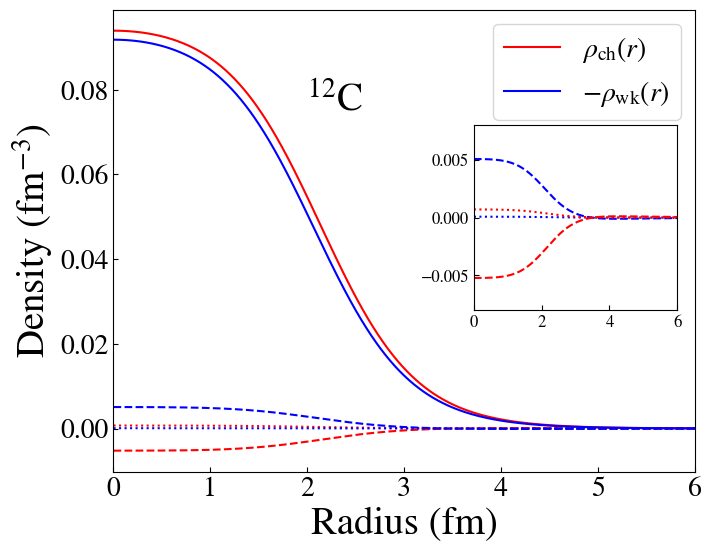}
    \caption{Densities for $^{12}$C, see \cref{fig:density}.}
    \label{fig:carbon_densities}
\end{figure}

\begin{table}[htbp]
\begin{center} 
\begin{tabular}{| l | c c | c c |}
\hline
 & \multicolumn{2}{c|}{$^{208}$Pb (1 GeV)} 
 & \multicolumn{2}{c|}{$^{48}$Ca (2.18 GeV)} \\
\cline{2-5}
Correction & 2nd B. (\%) & Eik. (\%) 
           & 2nd B. (\%) & Eik. (\%) \\
\hline
$v_{vs}$ & 4.19 & 4.73  & 5.88 & 6.29  \\
$v_{vs}^a$ & -4.07 & -3.66   & -5.77 & -5.55  \\
$v_{\pi}$ & -0.65 & -0.80  & -0.77 & -0.80 \\
$v_{vs}+v_{vs}^a$ & 0.12 & 0.89  & 0.11 & 0.38 \\
$v_{vs}+v_{\pi}$ & 3.54 & 3.86  & 5.11 & 5.39 \\
\cline{1-5}
$v_{vs}+v_{vs}^a+v_{\pi}$ & \bf{-0.53} & \bf{0.06}  & \bf{-0.66} & \bf{-0.47} \\
\hline
\end{tabular}
\end{center}
\caption{Radiative corrections as calculated in 2nd Born or eikonal for 1 GeV electrons scattering from $^{208}$Pb with $Q=0.4$ fm$^{-1}$ and 2.18 GeV electrons scattering from $^{48}$Ca with $Q=0.8733$ fm$^{-1}$. Last line gives the total correction.}
\label{tab:prex}
\end{table}

We first consider Born results for $^{208}$Pb at a momentum transfer of $Q=0.4$ fm$^{-1}$, see \cref{tab:prex} where we tabulate, 
\begin{equation}
  \Delta A_{\rm pv}=(A_{\rm pv}-A_{\rm pv}^b)/A_{\rm pv}^b\, ,  
\end{equation}
with $A_{\rm pv}$ either calculated in 2nd Born $A_{\rm pv}^{(2)}$ or Eikonal.  We find that indeed the Coulomb and axial vertex corrections have the largest impact on $A_{\rm pv}$ and have opposite signs in their impact.
As stated previously, they nearly cancel each other within a factor of $\alpha/2\pi$ at the 2nd Born level.
Further adding in the contribution from the vacuum polarization we see a decrease by about 0.5\% from the Born asymmetry without radiative corrections.
The same behavior can be seen in $^{48}$Ca at momentum transfer $Q=0.8733$ fm$^{-1}$ in \cref{tab:prex}.
We see a larger cancellation between the Coulomb and axial vertex corrections leaving only a relative change in $A_{\rm pv}$ in 2nd Born of approximately $0.66\%$.
This is because of the logarithmic behavior of the vacuum polarization in momentum space.

We also show our results for the two previous nuclei with the eikonal calculation of Coulomb distortions in the same tables.
Beginning with $^{208}$Pb we evaluate the eikonal calculation using a beam energy of $E=1$ GeV evaluated at the same $Q=0.4$ fm$^{-1}$, consistent with PREX 2 energetics \cite{PhysRevLett.126.172503}.
We find a noticeable change in all radiative corrections when including Coulomb distortions.
Firstly, the Coulomb vertex correction increases by 0.5\%, further increasing the Coulomb amplitude.
This, combined with the decrease in the axial vertex contribution, leads to a non-cancellation compared to the Born amplitude, leaving a leftover correction of almost 1\%.
Indeed, only after including the vacuum polarization which encounters a slight enhancement with Coulomb distortions, does the sum of corrections cancel leaving a sub 0.1\% total correction to $A_{\rm pv}$.
The reason for this is due to the short-range nature of the axial vertex compared to the long-range nature of the Coulomb vertex, wherein the Coulomb distortions further enhance the Coulomb vertex contribution to the scattering amplitude.

We see a much smaller change in $^{48}$Ca, where we use a beam energy of $E=2.18$ GeV evaluated at the same $Q=0.8733$ fm$^{-1}$, the quoted CREX kinematics \cite{PhysRevLett.129.042501}.
Being a much lighter nucleus than $^{208}$Pb, we observe a much smaller change from Born when calculating $A_{\rm pv}$ with Coulomb distortions.
We find the same trends as previously with $^{208}$Pb among the increases/decreases in the radiative corrections.
However, there is a more significant cancellation between the Coulomb and axial vertex corrections.
As such, the vacuum polarization brings the total radiative correction to approximately -0.5\% compared to no radiative corrections.
In turn, this \textit{decreases} the total asymmetry.

\begin{figure}[bth]
    \centering
    \includegraphics[width=1.\linewidth]{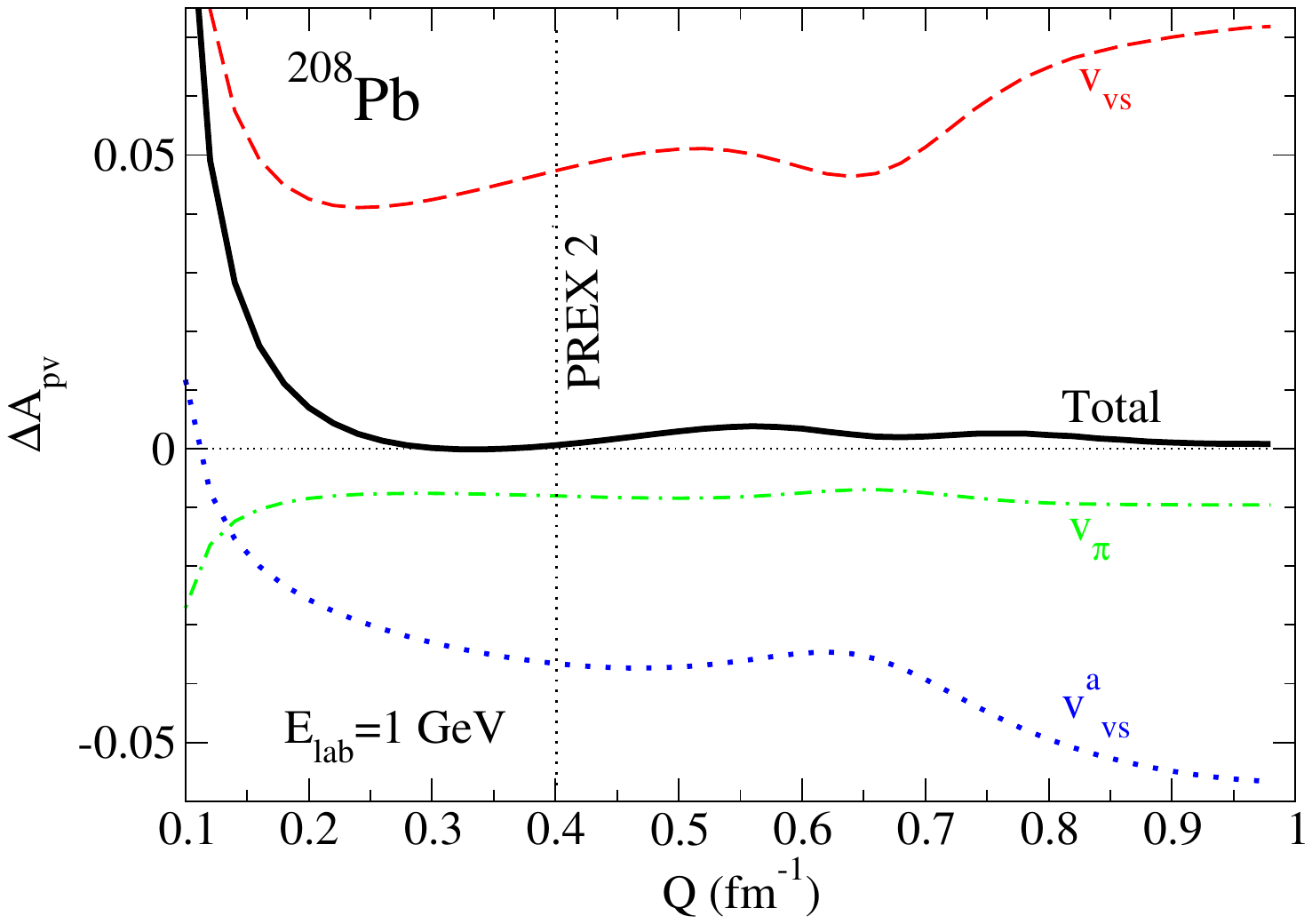}
    \caption{Radiative corrections for $^{208}$Pb at 1 GeV  calculated using Eikonal solutions. Vertical dashed line shows the central $Q$ value of PREX 2.}
    \label{fig:prex_eikonal}
\end{figure}

\begin{figure}[bth]
    \centering
    \includegraphics[width=1.\linewidth]{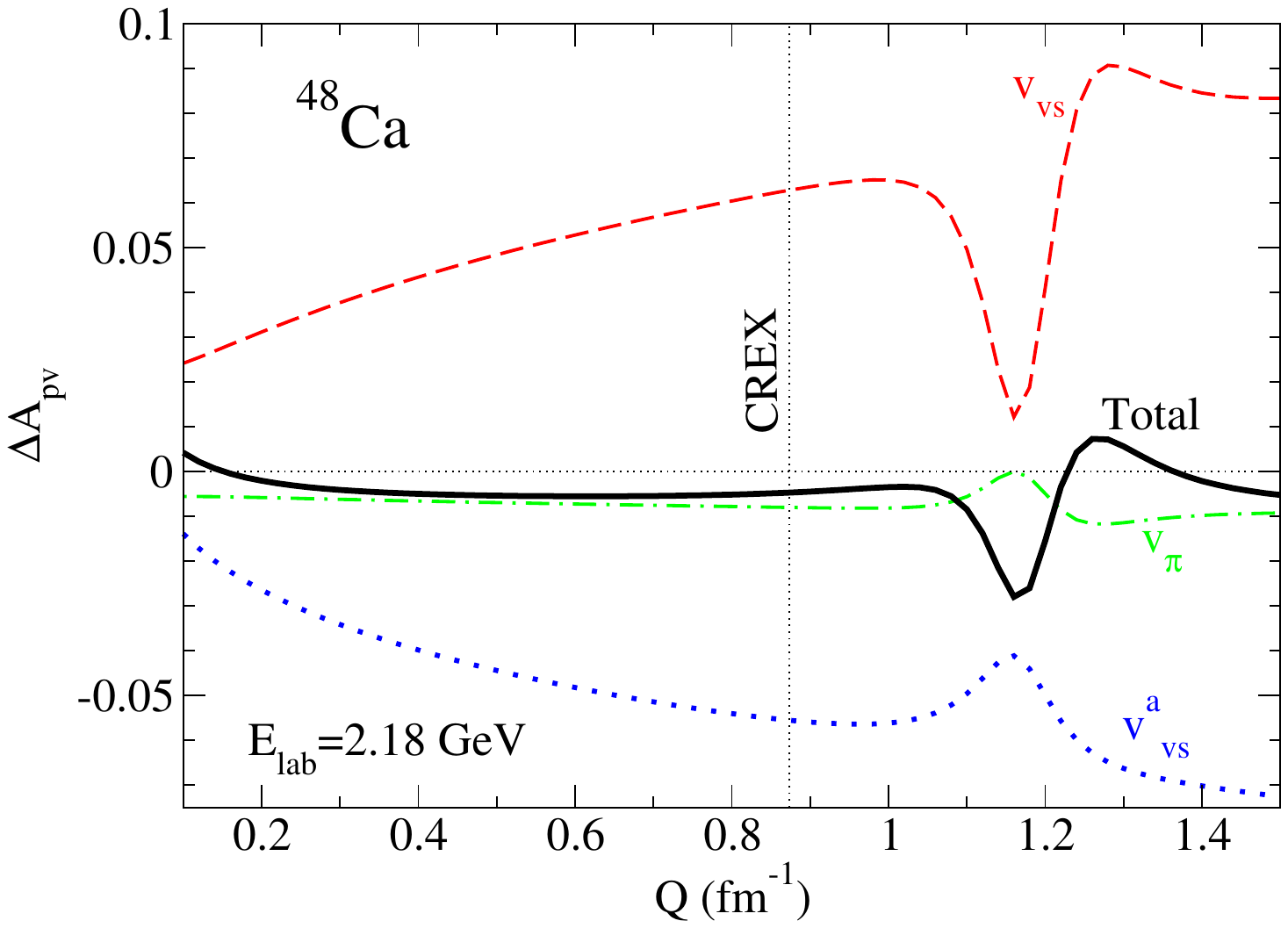}
    \caption{Radiative corrections for $^{48}$Ca at 2.18 GeV. Vertical dashed line shows the central $Q$ value of CREX.}
    \label{fig:crex_eikonal}
\end{figure}

\begin{figure}[htb]
    \centering
    \includegraphics[width=1\linewidth]{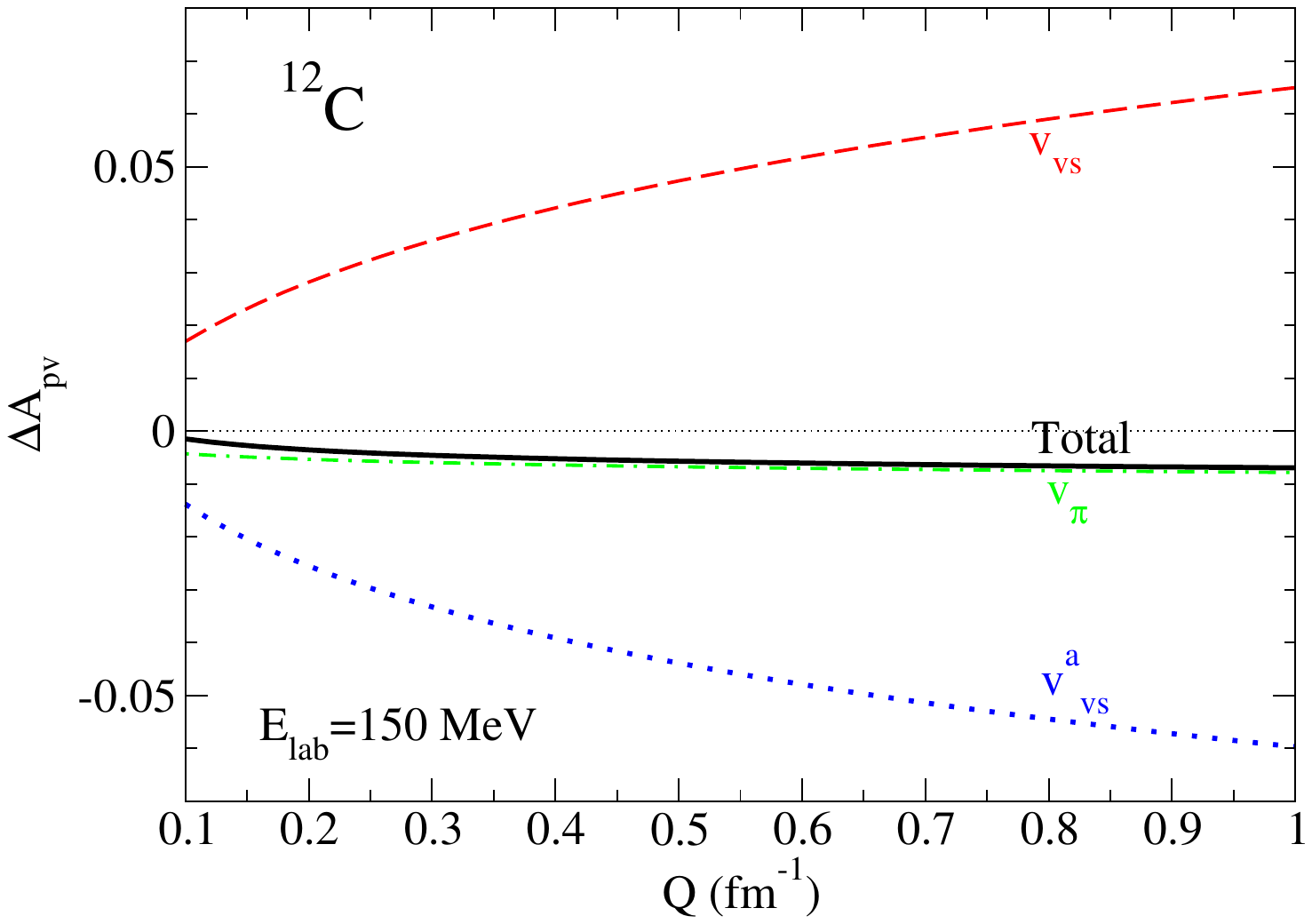}
    \caption{Radiative corrections for $^{12}$C at 150 MeV.}
    \label{fig:carbon_DA}
\end{figure}

To assess how these corrections to $A_{\rm pv}$ depend on momentum transfer, we show in \cref{fig:prex_eikonal,fig:crex_eikonal} the behavior of radiative corrections in $^{208}$Pb and $^{48}$Ca.
Firstly in $^{208}$Pb, we see that despite the vertex corrections increasing in strength as $Q$ increases, the overall effect by including vertex+vacuum polarization remains unchanged.
The small total correction (except at very small momentum transfer values) indicates that electroweak radiative corrections may be unimportant for PREX.
Similarly, we see the same behavior in $^{48}$Ca except for a noticeable dip in the vertex corrections near $Q\sim1.2$ fm$^{-1}$.
This is due to the diffraction minimum in the cross-section.  Outside of this range, however, we see that the sum of radiative corrections remains at around the -0.5\% value at all values of $Q$.  The -0.5\% total radiative correction is small compared to the 4\% statistical error of CREX \cite{PhysRevLett.129.042501}.

We conclude this section by investigating the possible implications of these corrections on future PVES experiments, namely parity-violating experiments at MESA.
Firstly, the Mainz Radius Experiment (MREX) is projected to provide a third measurement of the neutron skin in $^{208}$Pb at a much lower beam energy ($E\sim150$ MeV) with an expected precision improvement on PREX2 of a factor of two.
The radiative corrections presented here, however, are largely independent of beam energy.
As a result, the expected contribution from electroweak radiative correction to the interpretability of MREX is minimal.
Later, we discuss possible additional corrections which also need to be documented.

\begin{table}[htbp]
\begin{center} 
  \begin{tabular}{| l | r | r |}
    \hline
    
 Correction & 2nd Born (\%) & Eikonal (\%)  \\
    \hline
$v_{vs}$ & 4.19 & 4.22      \\
$v_{vs}^a$ & -4.07 & -3.91     \\
$v_{\pi}$ & -0.65& -0.64  \\
$v_{vs}+v_{vs}^a$ & 0.12 & 0.14  \\
$v_{vs}+v_{\pi}$& 3.54& 3.53\\
\hline
$v_{vs}+v_{vs}^a+v_{\pi}$ & \bf{-0.53} & \bf{-0.52}  \\\hline
    \end{tabular}
    \end{center}\caption{Radiative corrections for 150 MeV electrons scattering from $^{12}$C with $Q=0.4$ fm$^{-1}$ as calculated in 2nd Born and Eikonal.} \label{tab:carbon} 
\end{table}

MESA may also measure electron-scattering from $^{12}$C with a goal of determining $A_{\rm pv}$ and the weak charge of $^{12}$C to 0.3\%.
We show the radiative corrections for $^{12}$C in \cref{tab:carbon} and in \cref{fig:carbon_DA}.
Much more so than the two heavier nuclei discussed above, $^{12}$C shows very simple behavior in its radiative corrections.
In fact, the difference between the 2nd Born and the Coulomb-distorted corrections are very small $\mathcal{O}(0.1\%)$.
We see that the total radiative correction is of similar size as in $^{48}$Ca, at $\sim-0.5\%$.  This is larger than the experimental error goal of 0.3\%.  Therefore, radiative corrections must be carefully included.

\section{Discussion}
\label{sec:discussion}
We begin by discussing the cancellation between the axial-vector $v^a_{\rm vs}$ and vector $v_{\rm vs}$ vertices.  We are interested in on-shell spinor $u$ matrix elements of the vertices in \cref{vector_vertex,eq:Gamma_v} where the Dirac Eq. reads $mu=\slashed{k}u$.  Therefore, we can replace $m\gamma_\lambda$ with $\gamma_\lambda\slashed{k}$ in the numerators of \cref{vector_vertex,eq:Gamma_v} and then move $\gamma^5$ to the right to get,
\begin{equation}
    i\Gamma_v^\mu=e^2\int\frac{d^4l}{(2\pi)^4}\frac{\gamma^\lambda (\slashed k'+\slashed l+m)\gamma^\mu\big((\slashed k+\slashed l)\gamma_\lambda+\gamma_\lambda\slashed{k}\big)}{(l^2-m_\gamma^2)(l^2+2l\cdot k')(l^2+2l\cdot k)}\, .
\label{vector_vertex2}
\end{equation}
and,
\begin{equation}
    i\Gamma_a^\mu=e^2\int\frac{d^4l}{(2\pi)^4}\frac{\gamma^\lambda (\slashed k'+\slashed l+m)\gamma^\mu\big((\slashed k+\slashed l)\gamma_\lambda+\gamma_\lambda\slashed{k}\big)\gamma^5}{(l^2-m_\gamma^2)(l^2+2l\cdot k')(l^2+2l\cdot k)}.
\label{axial_vertex2}
\end{equation}
Now $\gamma^5u(k,h)=hu(k,h)$ for a spinor of helicity $h$ in the limit $k\gg m$.  Therefore, comparing \cref{vector_vertex2,axial_vertex2} gives 
\begin{equation}
\Gamma_a^\mu(Q^2)=\Gamma_v^\mu(Q^2)\gamma^5\, ,
\label{eq:va=vv}
\end{equation}
in the limit $Q^2\gg m^2$.  This explains why the $\ln Q^2$ and $\ln^2Q^2$ terms in $v_{\rm vs}$ in \cref{eq:vvs} and $v_{\rm vs}^a$ in \cref{eq:vavs} are identical. 

The only difference between $v_{\rm vs}^a$ and $v_{\rm vs}$ comes from different counter terms at $Q^2=0$.  By convention, the weak charge of the nucleus $Q_{\rm wk}$ includes radiative corrections defined at $Q^2=0$. Indeed, there may be a correction factor of $-\alpha/2\pi$ included in $Q_{\rm wk}$ from the axial vertex at $Q^2=0$.  However, parity violating electron scattering experiments are not performed at $Q^2=0$. Instead, they are typically performed at a modest $Q^2$ compared to hadronic scales. Nevertheless, the momentum transfer is still large compared to the small mass of the electron $Q^2\gg m^2$. Therefore, $v_{\rm vs}^a$ must be larger than $v_{\rm vs}$ by $\alpha/2\pi$, see \cref{eq:vavs-vvs}, to cancel the correction to $Q_{\rm wk}$ and satisfy \cref{eq:va=vv}.

We now discuss the overall impact of these radiative corrections on the interpretability of the PREX1+2 and CREX experiments.
It was previously suggested that E+M radiative corrections may be large \cite{PhysRevLett.134.192501}, resulting in largely unphysical interpretation of the main goal of these two campaigns: to measure the neutron-skin in these nuclei.
By only including the E+M results, we have shown that indeed one expects to find a large correction to $A_{\rm pv}$, mostly coming from the Coulomb vertex correction.
The cancellation that occurs between the axial and Coulomb corrections thus is incredibly important as the total correction we show is much smaller.
Following the analysis and extraction of the neutron-skin from both PREX2 and CREX, one finds an overall change to the extracted neutron skins to be less than 0.5\%, far below the quoted statistical and experimental uncertainties.
Thus, one major conclusion of our findings are that these electroweak radiative corrections presented here \textit{do not} impact the experimental results and implications of PREX1+2 nor CREX.  
Note that MREX is also projected to be largely unimpacted by such corrections.

\begin{figure}[htb]
    \centering
    \includegraphics[width=0.9\linewidth]{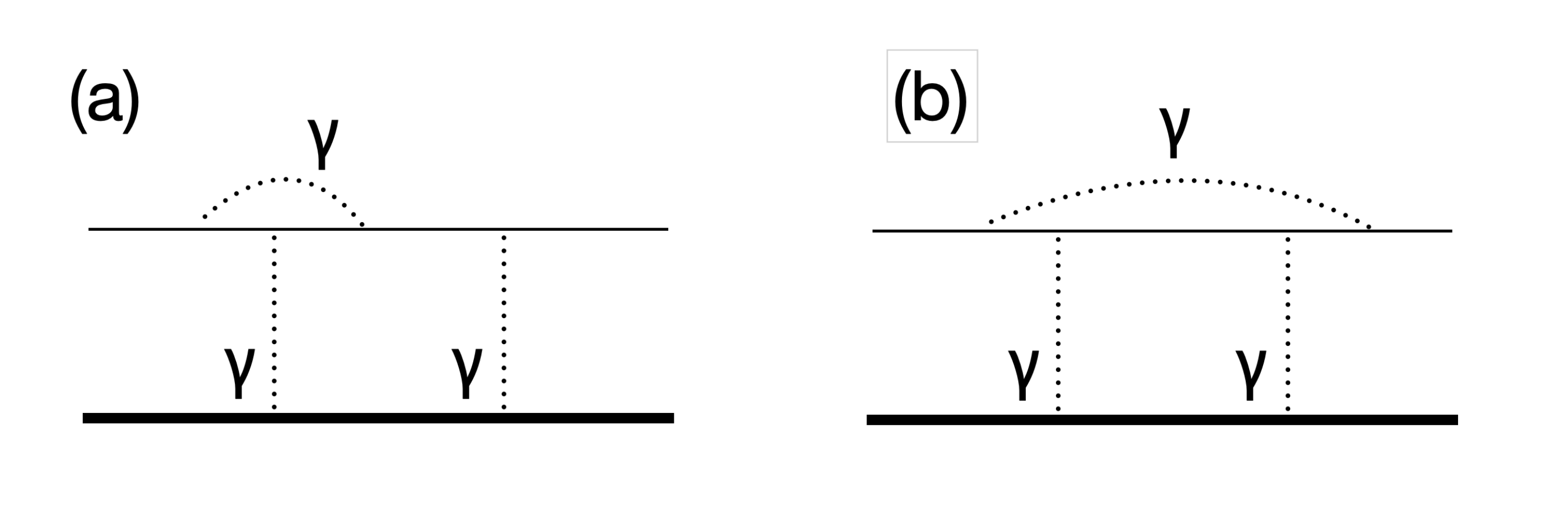}
    \caption{(a) Example of a Coulomb distortion correction to the vector vertex.  The electron line is thin while the heavy line shows the nucleus.   (b) Example of a diagram not included.}
    \label{fig:distortions}
\end{figure}

We note that our calculation of Coulomb distortions in \cref{sec:coulomb} is only approximate.  For the vector and axial-vector vertex corrections we include any number of additional Coulomb interactions before or after the vertex.  For example \cref{fig:distortions} (a) shows one Coulomb interaction after the vertex.  However, we do not include diagrams such as \cref{fig:distortions} (b) where there are additional Coulomb interactions inside the vertex.  These corrections remain to be examined.  However, the Coulomb effects we do include provide only modest changes to the radiative corrections.



It also remains to calculate corrections to $A_{\rm pv}$ from Bremsstrahlung.  
This depends on the energy resolution of the experiment and may be more complicated for the MREX experiment if the energy resolution is not much smaller than the beam energy.  
Bremsstrahlung results will be presented in future work.

\section{Conclusions}
\label{sec:conclusions}
In this paper we have calculated radiative corrections to the parity-violating asymmetry $A_{\rm pv}$ in electron-nucleus scattering including the vertex and vacuum polarization diagrams shown in \cref{fig:4diagrams}.  
We find large cancellations between the vertex corrections to the vector and axial-vector vertices.  
As a result the total correction in the 2nd Born approximation is dominated by vacuum polarization and is of order -0.5\%.  
Coulomb distortions have modest effects on radiative corrections for $^{12}$C and $^{48}$Ca nuclei.  
For $^{208}$Pb Coulomb distortions reduce the total radiative correction to only about 0.1\%.  
Therefore, these corrections are unimportant for the interpretation of the PREX, CREX, and MREX experiments. 
However radiative corrections must be carefully included for a precision measurement of the weak charge of $^{12}$C.

\begin{acknowledgments}
We would like to thank 
M. Vanderhaegen, C. Sfienti, E Mereghetti, \`Oscar Lara-Crosas, and Martin Hoferichter for helpful discussions regarding the calculation of radiative corrections and interpretation of the results.
Work supported in part by US Department of Energy grant DE-FG02-87ER40365, Laboratory Directed Research and Development Program of Los Alamos National Laboratory under project number 20230785PRD1, and National Science Foundation grant PHY 21-16686.  Los Alamos National Laboratory is operated by Triad National Security, LLC, for the National Nuclear Security Administration of U.S. Department of Energy (Contract No.~89233218CNA000001).

\end{acknowledgments}


\bibliography{references}

@article{PhysRevD.83.025020,
  title = {Gauge invariant treatment of ${\ensuremath{\gamma}}_{5}$ in the scheme of 't Hooft and Veltman},
  author = {Tsai, Er-Cheng},
  journal = {Phys. Rev. D},
  volume = {83},
  issue = {2},
  pages = {025020},
  numpages = {16},
  year = {2011},
  month = {Jan},
  publisher = {American Physical Society},
  doi = {10.1103/PhysRevD.83.025020},
  url = {https://link.aps.org/doi/10.1103/PhysRevD.83.025020}
}

@article{PhysRevC.102.022501,
  title = {Weak charge and weak radius of $^{12}\mathrm{C}$},
  author = {Koshchii, Oleksandr and Erler, Jens and Gorchtein, Mikhail and Horowitz, Charles J. and Piekarewicz, Jorge and Roca-Maza, Xavier and Seng, Chien-Yeah and Spiesberger, Hubert},
  journal = {Phys. Rev. C},
  volume = {102},
  issue = {2},
  pages = {022501},
  numpages = {6},
  year = {2020},
  month = {Aug},
  publisher = {American Physical Society},
  doi = {10.1103/PhysRevC.102.022501},
  url = {https://link.aps.org/doi/10.1103/PhysRevC.102.022501}
}

@misc{schlimme2024mesaphysicsprogram,
      title={The MESA physics program}, 
      author={Sören Schlimme and Kurt Aulenbacher and Sebastian Baunack and Niklaus Berger and Achim Denig and Luca Doria and Alfons Khoukaz and Frank Maas and Harald Merkel and Concettina Sfienti and Michaela Thiel},
      year={2024},
      eprint={2402.01027},
      archivePrefix={arXiv},
      primaryClass={nucl-ex},
      url={https://arxiv.org/abs/2402.01027}, 
}

@article{PhysRevLett.86.5647,
  title = {Neutron Star Structure and the Neutron Radius of $^{208}Pb$},
  author = {Horowitz, C. J. and Piekarewicz, J.},
  journal = {Phys. Rev. Lett.},
  volume = {86},
  issue = {25},
  pages = {5647--5650},
  numpages = {0},
  year = {2001},
  month = {Jun},
  publisher = {American Physical Society},
  doi = {10.1103/PhysRevLett.86.5647},
  url = {https://link.aps.org/doi/10.1103/PhysRevLett.86.5647}
}

@article{PhysRevLett.95.122501,
  title = {Neutron-Rich Nuclei and Neutron Stars: A New Accurately Calibrated Interaction for the Study of Neutron-Rich Matter},
  author = {Todd-Rutel, B. G. and Piekarewicz, J.},
  journal = {Phys. Rev. Lett.},
  volume = {95},
  issue = {12},
  pages = {122501},
  numpages = {4},
  year = {2005},
  month = {Sep},
  publisher = {American Physical Society},
  doi = {10.1103/PhysRevLett.95.122501},
  url = {https://link.aps.org/doi/10.1103/PhysRevLett.95.122501}
}

@article{PhysRevLett.128.132501,
  title = {Determination of the $^{27}\mathrm{Al}$ Neutron Distribution Radius from a Parity-Violating Electron Scattering Measurement},
  author = {Androi\ifmmode \acute{c}\else \'{c}\fi{}, D. and Armstrong, D. S. and Bartlett, K. and Beminiwattha, R. S. and Benesch, J. and Benmokhtar, F. and Birchall, J. and Carlini, R. D. and Cornejo, J. C. and Covrig Dusa, S. and others},
  collaboration = {${\mathrm{Q}}_{\mathrm{weak}}$ Collaboration},
  journal = {Phys. Rev. Lett.},
  volume = {128},
  issue = {13},
  pages = {132501},
  numpages = {7},
  year = {2022},
  month = {Apr},
  publisher = {American Physical Society},
  doi = {10.1103/PhysRevLett.128.132501},
  url = {https://link.aps.org/doi/10.1103/PhysRevLett.128.132501}
}

@article{PhysRevLett.126.172503,
  title = {Implications of PREX-2 on the Equation of State of Neutron-Rich Matter},
  author = {Reed, Brendan T. and Fattoyev, F. J. and Horowitz, C. J. and Piekarewicz, J.},
  journal = {Phys. Rev. Lett.},
  volume = {126},
  issue = {17},
  pages = {172503},
  numpages = {5},
  year = {2021},
  month = {Apr},
  publisher = {American Physical Society},
  doi = {10.1103/PhysRevLett.126.172503},
  url = {https://link.aps.org/doi/10.1103/PhysRevLett.126.172503}
}

@article{PhysRevLett.108.112502,
  title = {Measurement of the Neutron Radius of $^{208}\mathrm{Pb}$ through Parity Violation in Electron Scattering},
  author = {Abrahamyan, S. and Ahmed, Z. and Albataineh, H. and Aniol, K. and Armstrong, D. S. and Armstrong, W. and Averett, T. and Babineau, B. and Barbieri, A. and Bellini, V. and 
  others},
  collaboration = {PREX Collaboration},
  journal = {Phys. Rev. Lett.},
  volume = {108},
  issue = {11},
  pages = {112502},
  numpages = {6},
  year = {2012},
  month = {Mar},
  publisher = {American Physical Society},
  doi = {10.1103/PhysRevLett.108.112502},
  url = {https://link.aps.org/doi/10.1103/PhysRevLett.108.112502}
}

@article{PhysRevC.102.044321,
  title = {Insights into nuclear saturation density from parity-violating electron scattering},
  author = {Horowitz, C. J. and Piekarewicz, J. and Reed, Brendan},
  journal = {Phys. Rev. C},
  volume = {102},
  issue = {4},
  pages = {044321},
  numpages = {6},
  year = {2020},
  month = {Oct},
  publisher = {American Physical Society},
  doi = {10.1103/PhysRevC.102.044321},
  url = {https://link.aps.org/doi/10.1103/PhysRevC.102.044321}
}

@article{Thiel_2019,
doi = {10.1088/1361-6471/ab2c6d},
url = {https://doi.org/10.1088/1361-6471/ab2c6d},
year = {2019},
month = {aug},
publisher = {IOP Publishing},
volume = {46},
number = {9},
pages = {093003},
author = {Thiel, M and Sfienti, C and Piekarewicz, J and Horowitz, C J and Vanderhaeghen, M},
title = {Neutron skins of atomic nuclei: per aspera ad astra},
journal = {Journal of Physics G: Nuclear and Particle Physics}
}

@article{DONNELLY1989589,
title = {Isospin dependences in parity-violating electron scattering},
journal = {Nuclear Physics A},
volume = {503},
number = {3},
pages = {589-631},
year = {1989},
issn = {0375-9474},
doi = {https://doi.org/10.1016/0375-9474(89)90432-6},
url = {https://www.sciencedirect.com/science/article/pii/0375947489904326},
author = {T.W. Donnelly and J. Dubach and Ingo Sick}
}

@article{PhysRevC.63.025501,
  title = {Parity violating measurements of neutron densities},
  author = {Horowitz, C. J. and Pollock, S. J. and Souder, P. A. and Michaels, R.},
  journal = {Phys. Rev. C},
  volume = {63},
  issue = {2},
  pages = {025501},
  numpages = {18},
  year = {2001},
  month = {Jan},
  publisher = {American Physical Society},
  doi = {10.1103/PhysRevC.63.025501},
  url = {https://link.aps.org/doi/10.1103/PhysRevC.63.025501}
}

@article{PhysRevLett.129.042501,
  title = {Precision Determination of the Neutral Weak Form Factor of $^{48}\mathrm{Ca}$},
  author = {Adhikari, D. and Albataineh, H. and Androic, D. and Aniol, K. A. and Armstrong, D. S. and Averett, T. and Ayerbe Gayoso, C. and Barcus, S. K. and Bellini, V. and Beminiwattha, R. S. and others },
  collaboration = {CREX Collaboration},
  journal = {Phys. Rev. Lett.},
  volume = {129},
  issue = {4},
  pages = {042501},
  numpages = {8},
  year = {2022},
  month = {Jul},
  publisher = {American Physical Society},
  doi = {10.1103/PhysRevLett.129.042501},
  url = {https://link.aps.org/doi/10.1103/PhysRevLett.129.042501}
}

@article{PhysRevLett.126.172502,
  title = {Accurate Determination of the Neutron Skin Thickness of $^{208}\mathrm{Pb}$ through Parity-Violation in Electron Scattering},
  author = {Adhikari, D. and Albataineh, H. and Androic, D. and Aniol, K. and Armstrong, D. S. and Averett, T. and Ayerbe Gayoso, C. and Barcus, S. and Bellini, V. and Beminiwattha, R. S. and others
  },
  collaboration = {PREX Collaboration},
  journal = {Phys. Rev. Lett.},
  volume = {126},
  issue = {17},
  pages = {172502},
  numpages = {7},
  year = {2021},
  month = {Apr},
  publisher = {American Physical Society},
  doi = {10.1103/PhysRevLett.126.172502},
  url = {https://link.aps.org/doi/10.1103/PhysRevLett.126.172502}
}

@article{PhysRevLett.134.192501,
  title = {QED Corrections to the Parity-Violating Asymmetry in High-Energy Electron-Nucleus Collisions},
  author = {Roca-Maza, Xavier and Jakubassa-Amundsen, D. H.},
  journal = {Phys. Rev. Lett.},
  volume = {134},
  issue = {19},
  pages = {192501},
  numpages = {5},
  year = {2025},
  month = {May},
  publisher = {American Physical Society},
  doi = {10.1103/PhysRevLett.134.192501},
  url = {https://link.aps.org/doi/10.1103/PhysRevLett.134.192501}
}

@ARTICLE{Vanderhaeghen:2000,
       author = {{Vanderhaeghen}, M. and {Friedrich}, J.~M. and {Lhuillier}, D. and {Marchand}, D. and {van Hoorebeke}, L. and {van de Wiele}, J.},
        title = "{QED radiative corrections to virtual Compton scattering}",
      journal = {\prc},
         year = 2000,
        month = aug,
       volume = {62},
       number = {2},
          eid = {025501},
        pages = {025501},
          doi = {10.1103/PhysRevC.62.025501},
archivePrefix = {arXiv},
       eprint = {hep-ph/0001100},
 primaryClass = {hep-ph},
       adsurl = {https://ui.adsabs.harvard.edu/abs/2000PhRvC..62b5501V}
}

@BOOK{qft_book,
       author = {{Peskin}, Michael E. and {Schroeder}, Daniel V.},
        title = "{An Introduction to Quantum Field Theory}",
         year = 1995,
       adsurl = {https://ui.adsabs.harvard.edu/abs/1995iqft.book.....P}
}

@ARTICLE{milstein:2005,
       author = {{Milstein}, A.~I. and {Sushkov}, O.~P.},
        title = "{Vacuum polarization radiative correction to parity violating electron scattering on heavy nuclei}",
      journal = {\prc},
         year = 2005,
        month = apr,
       volume = {71},
       number = {4},
          eid = {045503},
        pages = {045503},
          doi = {10.1103/PhysRevC.71.045503},
archivePrefix = {arXiv},
       eprint = {hep-ph/0409149},
 primaryClass = {hep-ph},
       adsurl = {https://ui.adsabs.harvard.edu/abs/2005PhRvC..71d5503M}
}

@article{Reed_2020st,
	title        = {Measuring the surface thickness of the weak charge density of nuclei},
	author       = {Brendan T. Reed and Z. Jaffe and C. J. Horowitz and C. Sfienti},
	year         = 2020,
	month        = {dec},
	journal      = {Physical Review C},
	publisher    = {American Physical Society ({APS})},
	volume       = 102,
	number       = 6,
	pages        = {064308},
	doi          = {10.1103/physrevc.102.064308},
	url          = {https://doi.org/10.1103\%2Fphysrevc.102.064308},
	eprint       = {2009.06664},
	archiveprefix = {arXiv},
	primaryclass- = {nucl-th}
}

@article{Horowitz:1998vv,
	title        = {{Parity violating elastic electron scattering and Coulomb distortions}},
	author       = {Horowitz, C.J.},
	year         = 1998,
	journal      = {Phys. Rev. },
	volume       = {C57},
	pages        = {3430--3436},
	doi          = {10.1103/PhysRevC.57.3430},
	eprint       = {nucl-th/9801011},
	archiveprefix = {arXiv},
	primaryclass- = {nucl-th},
	reportnumber = {IU-NTC-97-12},
	slaccitation = {%%CITATION = NUCL-TH/9801011;%%}
}

@article{Piekarewicz:2016vbn,
	title        = {{Power of two: Assessing the impact of a second measurement of the weak-charge form factor of $^{208}$Pb}},
	author       = {Piekarewicz, J. and Linero, A. R. and Giuliani, P. and Chicken, E.},
	year         = 2016,
	journal      = {Phys. Rev.},
	volume       = {C94},
	number       = 3,
	pages        = {034316},
	doi          = {10.1103/PhysRevC.94.034316},
	slaccitation = {%%CITATION = ARXIV:1604.07799;%%},
	eprint       = {1604.07799},
	archiveprefix = {arXiv},
	primaryclass- = {nucl-th}
}

@article{CREX,
  title = {Precision Determination of the Neutral Weak Form Factor of $^{48}\mathrm{Ca}$},
  author = {Adhikari, D. and others},
  collaboration = {CREX Collaboration},
  journal = {Phys. Rev. Lett.},
  volume = {129},
  issue = {4},
  pages = {042501},
  numpages = {8},
  year = {2022},
  month = {Jul},
  publisher = {American Physical Society},
  doi = {10.1103/PhysRevLett.129.042501},
  url = {https://link.aps.org/doi/10.1103/PhysRevLett.129.042501}
}

@preamble{
 "\providecommand{\noopsort}[1]{}" 
 # "\providecommand{\singleletter}[1]{#1}%" 
}

@article{dong:2008,
	title        = {Relativistic eikonal approaches to parity violating electron-nucleus scattering},
	author       = {Dong, Tiekuang and Ren, Zhongzhou and Wang, Zaijun},
	year         = 2008,
	month        = {Jun},
	journal      = {Phys. Rev. C},
	publisher    = {American Physical Society},
	volume       = 77,
	doi          = {10.1103/PhysRevC.77.064302},
	url          = {https://link.aps.org/doi/10.1103/PhysRevC.77.064302},
	issue        = 6,
	numpages     = 10
}

@PREAMBLE{
 "\providecommand{\noopsort}[1]{}" 
 # "\providecommand{\singleletter}[1]{#1}%" 
}

\end{document}